# Top Mass from Electroweak Measurements[1]

Bob Jacobsen

CERN, PPE Division
CH-1211 Geneva 23
Switzerland

The electroweak measurements made at LEP using 1989-1993 data are presented in preliminary form. The agreement with the Standard Model is satisfactory, and allows a combined fit to all available data for the masses of the top quark and standard Higgs boson. The fit yields $M_t = 177^{+11+18}_{-11-19}$ GeV/c$^2$, where the second error reflects the uncertainty in the Higgs mass.

## 1  Introduction

The 1993 LEP running was dedicated to a scan of the Z boson lineshape[1]. The errors on measurements of other electroweak[2] observables, notably the asymmetries[3] and heavy quark widths, were also improved. Table 1 lists the 1993 preliminary combined LEP values. This talk presented these measurements in light of what they tell us about the mass of the top quark in the minimal Standard Model.[4]

The combination of an approximately four-fold increase in off-peak data, and an improved calibration[5] of the LEP energy, allowed the combined error from the four LEP experiments on the Z boson mass to be decreased from 7 to 4.4 MeV/c$^2$, and on the width from 7 to 3.8 MeV/c$^2$. The mass error is now dominated by the systematic uncertainty in the LEP energy, while the width error has approximately equal contributions from statistics and LEP energy uncertainty. Small additional improvements are expected when the measurements are final. This has resulted in significant improvements on the leptonic branching ratio $R_l$. Improvements in several of the LEP experiment's luminosity calorimeters have significantly reduced the systematic errors on absolute cross section measurements, leading to improved errors on the peak hadronic cross-section, $\sigma_h^0$.

---

[1]Talk presented at the XXIXth Rencontres de Moriond, "QCD and Hadronic Interactions", Meribel, France, March 19-26, 1994

|  | Measurement | Standard Model fit value | pull |
|---|---|---|---|
| **LEP** | | | |
| line shape: | | | |
| $M_Z$ (GeV) | 91.1895 ± 0.044 | 91.192 | 0.6 |
| $\Gamma_Z$ (GeV) | 2.4969 ± 0.0038 | 2.4967 | 0.1 |
| $\sigma_h^0$ (nb) | 41.51 ± 0.12 | 41.44 | 0.6 |
| $R_l$ | 20.789 ± 0.040 | 20.781 | 0.2 |
| $A_{FB}^{0,l}$ | 0.0170 ± 0.0016 | 0.0152 | 1.1 |
| $\tau$ polarization: | | | |
| $A_\tau$ | 0.150 ± 0.010 | 0.142 | 0.8 |
| $A_e$ | 0.120 ± 0.012 | 0.142 | 1.8 |
| b and c quark results: | | | |
| $R_b \equiv \Gamma_{b\bar{b}}/\Gamma_{had}$ | 0.2208 ± 0.0024 | 0.2158 | 2.0 |
| $R_c \equiv \Gamma_{c\bar{c}}/\Gamma_{had}$ | 0.170 ± 0.014 | 0.172 | 0.1 |
| $A_{FB}^{0,b}$ | 0.0960 ± 0.0043 | 0.0997 | 0.8 |
| $A_{FB}^{0,c}$ | 0.070 ± 0.011 | 0.071 | 0.1 |
| $q\bar{q}$ charge asymmetry: | | | |
| $\sin^2\theta_W^{eff}$ from $\langle Q_{FB}\rangle$ | 0.2320 ± 0.0016 | 0.2321 | 0.1 |
| **$p\bar{p}$ and $\nu$N** | | | |
| $M_W$ (GeV) | 80.23 ± 0.18 | 80.31 | 0.4 |
| $1 - M_W^2/M_Z^2$ ($\nu$N) | 0.2256 ± 0.0047 | 0.2246 | 0.2 |
| **SLC** | | | |
| $\sin^2\theta_W^{eff}$ from $A_e$ | 0.2294 ± 0.0010 | 0.2321 | 2.7 |

Table 1: Data input and results of the combined Standard Model fit. All LEP values are preliminary. The $p\bar{p}$ collider data are from UA2[6] CDF[7][8] and D0[8]. The neutrino experiment data are from CDHS[9] CHARM[10] and CCFR[11]. The SLC value is from the SLD measurement[12] of the left-right asymmetry. The second column is the measurement value and error used in the combined fits. The third column value results from the fit to all data (fourth column of Table 3), and the fourth column is the difference between the measured and fit values, normalized by the measurement error.

All asymmetries at the Z can be used to measure the weak coupling angle $\sin^2\theta_W^{eff}$. The measurements of lepton and c quark asymmetries, the polarization of tau leptons, and the total charge asymmetry have also been improved from 1992 values. The most precise estimate comes from the forward-backward asymmetry of b quark events (See Table 2). Combining these to a single estimate of $\sin^2\theta_W^{eff}$ from LEP data alone gives 0.2322±0.005 with a $\chi^2$/d.o.f of 6.3/5, corresponding to a 28% confidence level. Including the SLD measurement of $\sin^2\theta_W^{eff}$ from the left-right asymmetry increases the $\chi^2$/d.o.f to 12.8/6, giving a confidence level of 5%. The biggest difference in the combined set of measurements is between the values of $\sin^2\theta_W^{eff}$ from $A_{LR}$ and the $\tau$ forward-backward asymmetry, which both measure the electron coupling $A_e$, and are dominated by statistical errors.

|  | Measurement | Implied $\sin^2\theta_W^{eff}$ value |
|---|---|---|
| **LEP** | | |
| $A_{FB}^{0,l}$ | 0.0170 ± 0.0016 | 0.2311 ± 0.0009 |
| $A_\tau$ | 0.150 ± 0.010 | 0.2311 ± 0.0013 |
| $A_e$ | 0.120 ± 0.012 | 0.2350 ± 0.0015 |
| $A_{FB}^{0,b}$ | 0.0960 ± 0.0043 | 0.2328 ± 0.0008 |
| $A_{FB}^{0,c}$ | 0.070 ± 0.011 | 0.2324 ± 0.0026 |
| $\sin^2\theta_W^{eff}$ from $\langle Q_{FB}\rangle$ | 0.2320 ± 0.0016 | 0.2320 ± 0.0016 |
| **SLC** | | |
| $A_{LR}$ | 0.1637 ± 0.0075 | 0.2294 ± 0.0010 |

Table 2: $\sin^2\theta_W^{eff}$ central values and errors implied by various measurements. LEP numbers are averages of the four experiment's values.

The Z partial widths to b and c quarks have improved errors, due primarily to the use of new lifetime-based techniques. These measurements are now dominated by systematic errors, but all four experiments expect further progress in the coming year.

## 2  Standard Model Fits

The values listed in Table 1 were used as input to a Standard Model fit, and the results are summarized in Table 3. Assuming a Higgs mass of 300 GeV/c$^2$, the top mass is fitted to be 177±11 GeV/c$^2$ using all data. The improvement since the 1993 summer conferences is primarily due to the new Z width measurement, and the measurement is dominated by the LEP data. The $\alpha_s(M_Z^2)$ value implied by the fit is 0.124 ± 0.005 ± 0.002, quite consistent with independent measurements at LEP and elsewhere.

|  | LEP only | LEP, collider and ν data | LEP, SLC, collider and ν data |
|---|---|---|---|
| $M_t$ (GeV) | $172^{+13+18}_{-14-20}$ | $170^{+12+18}_{-12-20}$ | $177^{+11+18}_{-11-19}$ |
| $\alpha_s(M_Z^2)$ | $0.125 \pm 0.005 \pm 0.002$ | $0.125 \pm 0.005 \pm 0.002$ | $0.124 \pm 0.005 \pm 0.002$ |
| $\chi^2$/d.o.f | 11.4/9 | 11.5/11 | 19.1/12 |
| $\sin^2\theta_W^{eff}$ | $0.2323 \pm 0.0002^{+0.0001}_{-0.0002}$ | $0.2324 \pm 0.0002^{+0.0001}_{-0.0002}$ | $0.2320 \pm 0.0003^{+0.0001}_{-0.0002}$ |
| $1 - M_W^2/M_Z^2$ | $0.2251 \pm 0.0015^{+0.0003}_{-0.0003}$ | $0.2253 \pm 0.0013^{+0.0003}_{-0.0002}$ | $0.2243 \pm 0.0012^{+0.0003}_{-0.0002}$ |
| $M_W$ (GeV) | $80.28 \pm 0.08^{+0.01}_{-0.02}$ | $80.26 \pm 0.07^{+0.01}_{-0.01}$ | $80.31 \pm 0.06^{+0.01}_{-0.01}$ |

Table 3: Results of fits within the Standard Model hypothesis of the data of Table 1. The central values and first errors are quoted for a fixed Higgs mass of 300 GeV/c². The second error reflects the change in central value when the Higgs mass is varied from 60 GeV/c² to 1000 GeV/c². Column two includes only the LEP data (top section in Table 1), column three includes the $p\bar{p}$ and νN data from the center section of Table 1, and column four also includes the SLD result from the left-right asymmetry. $\alpha_s(M_Z^2)$ has been left free in all fits.

The $R_b$ and the forward-backward τ polarization asymmetry measurements contribute 4.0 and 3.2 respectively to the $\chi^2$ and tend to reduce the fitted top mass. The SLD measurement of $A_{LR}$ tends to increase the top mass and contributes 7.3 to the $\chi^2$. The $\chi^2$/d.o.f corresponds to a confidence level for the Standard Model hypothesis of 8.6% for the fit to all data, which is certainly acceptable. The fit to just LEP data and LEP plus collider and neutrino data have confidence levels of 28% and 40% respectively.

$\Gamma_{lepton}$, $\sin^2\theta_W^{eff}$, and $R_b$ give approximately independent constraints on $M_t$. Figure 1 shows the $\Gamma_{lepton}$ vs. $\sin^2\theta_W^{eff}$ plane with the Standard Model predictions and combined fit results overlaid. The preferred top and Higgs masses are clearly correlated in this projection.

As can be seen in Figure 2, the Standard Model value for $R_b$ is almost independent of the Higgs mass[13]. The measured $R_b$ value alone provides a 90% confidence level upper limit on the top mass of 180 GeV/c². Figure 3 is a different way of plotting the same data and shows the effect of $R_b$, when combined with the $R_{had}$ and the combined $\sin^2\theta_W^{eff}$ measurements, is to prefer lower Higgs masses. Figure 4 shows the variation of the fit $\chi^2$ as a function of top mass for three Higgs mass assumptions. The variation with Higgs mass is not large enough to be considered statistically significant, so the entire range of 60 to 1000 GeV/c² is used to estimate the uncertainty in the top mass due to the unknown Higgs mass.

The final value and error for the top mass is then $M_t = 177^{+11+18}_{-11-19}$ GeV/c², where the second error represents the uncertainty due to the unknown Higgs mass.

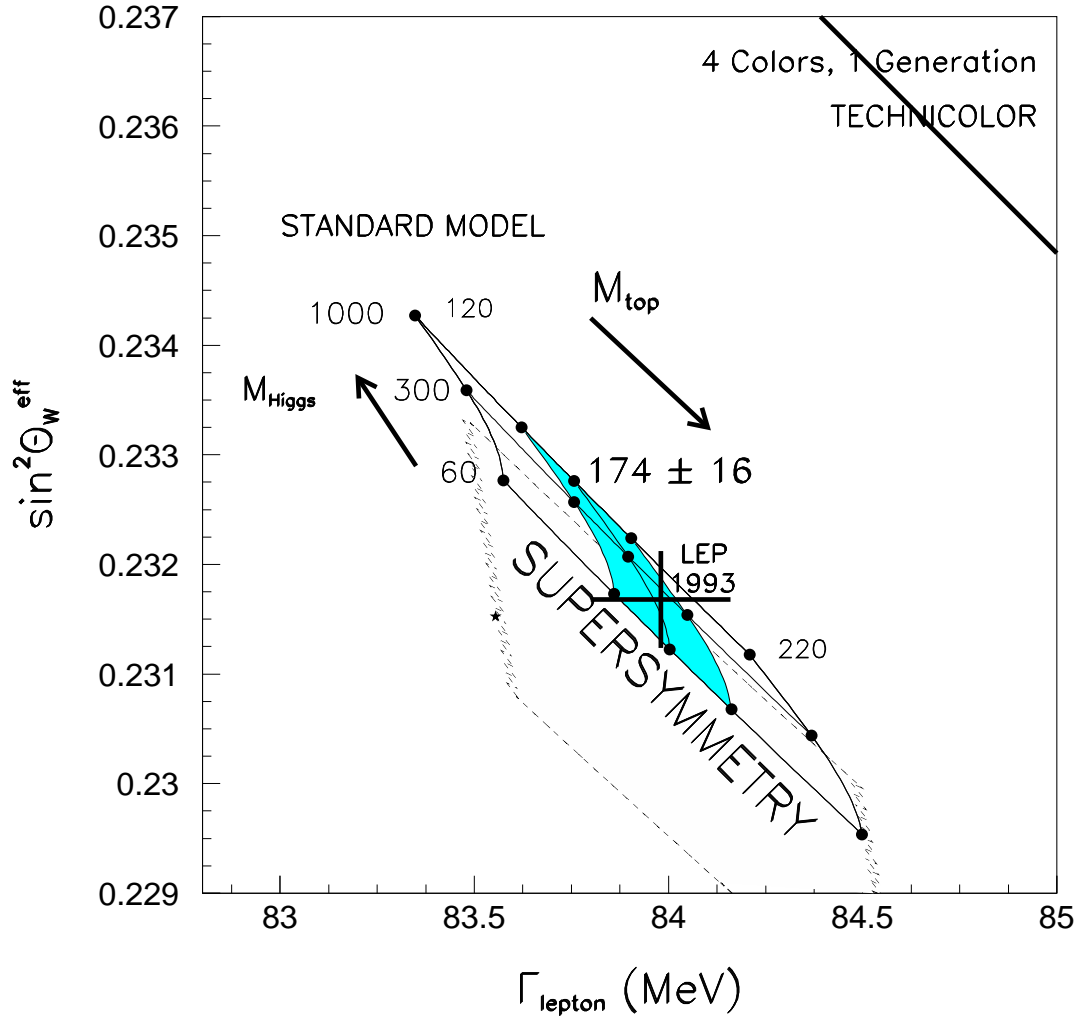

Figure 1: The $\sin^2\theta_W^{\text{eff}}$ vs. $\Gamma_{\text{lepton}}$ plane, overlaid with the results of the fit to all data and Standard Model predictions. The shaded band is the 1 σ limit on the top mass from CDF. The ★ indicates the Standard Model expectation without electroweak radiative corrections.

## Acknowledgments

The LEP results presented here were contributed by the LEP collaborations. The averaging and combined fits were done by the LEP Electroweak Working Group. I would particularly like to thank A. Blondel, P. Clarke, J. Harton, M. Koratzinos, M. Martinez, R. Miquel, M. Pepe-Alterelli, B. Pietrzyk, D. Schaile, and R. Tenchini for their help.

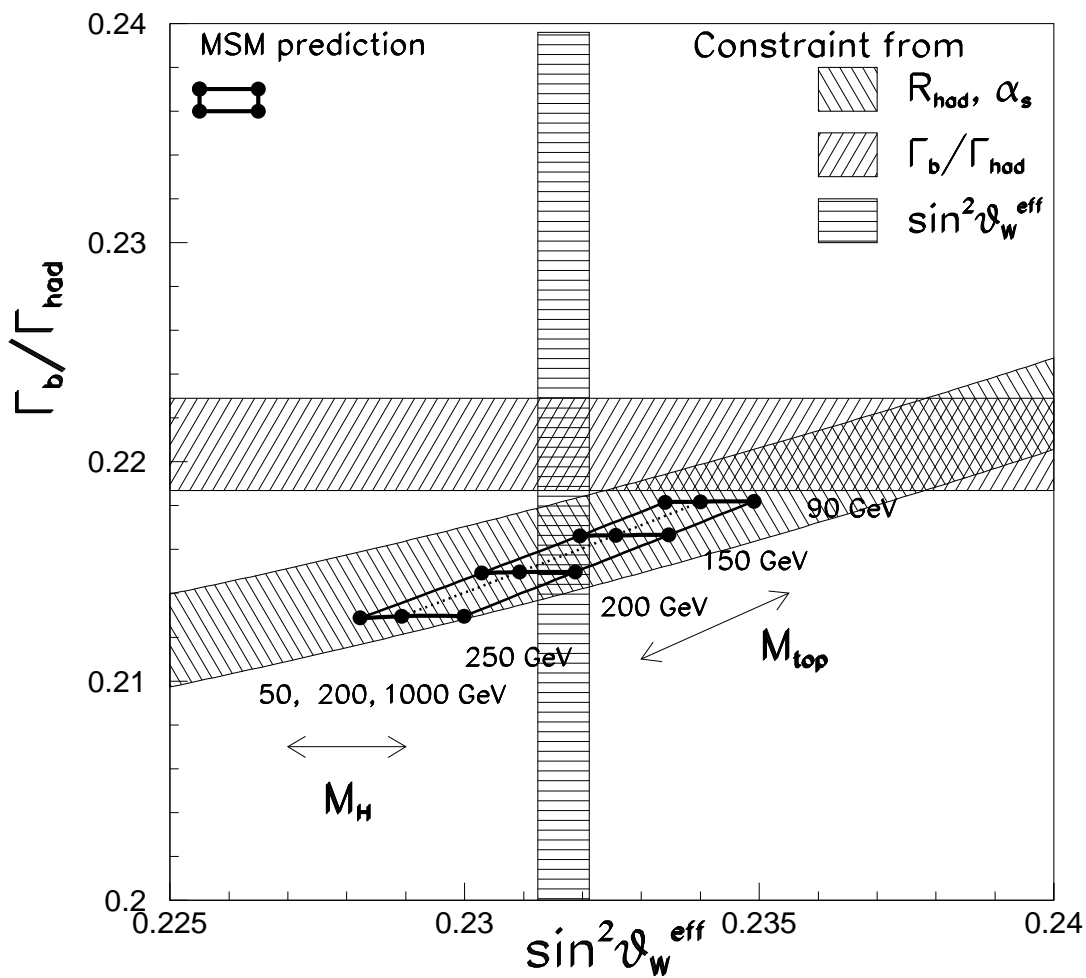

Figure 2: The $R_b$ vs. $\sin^2\theta_W^{eff}$ plane, overlaid with current measurements and Standard Model predictions. An $\alpha_s(M_Z^2)$ value of $0.123\pm0.006$ was used to produce this plot; lower values would move the $R_{had}$ band up.

# References

[1] P. Clarke, presented at the XXIXth Rencontres de Moriond, March 12-19 1994

[2] CDF's evidence for a top quark mass of $174\pm10\pm13$ was presented after this talk was delivered, and is not included in these results. Their mass estimate is consistent with the LEP predictions F. Abe et.al., Fermilab-Pub-94/097-E, submitted to Phys. Rev. D.

[3] B. Pietrzyk, presented at the XXIXth Rencontres de Moriond, March 12-19 1994

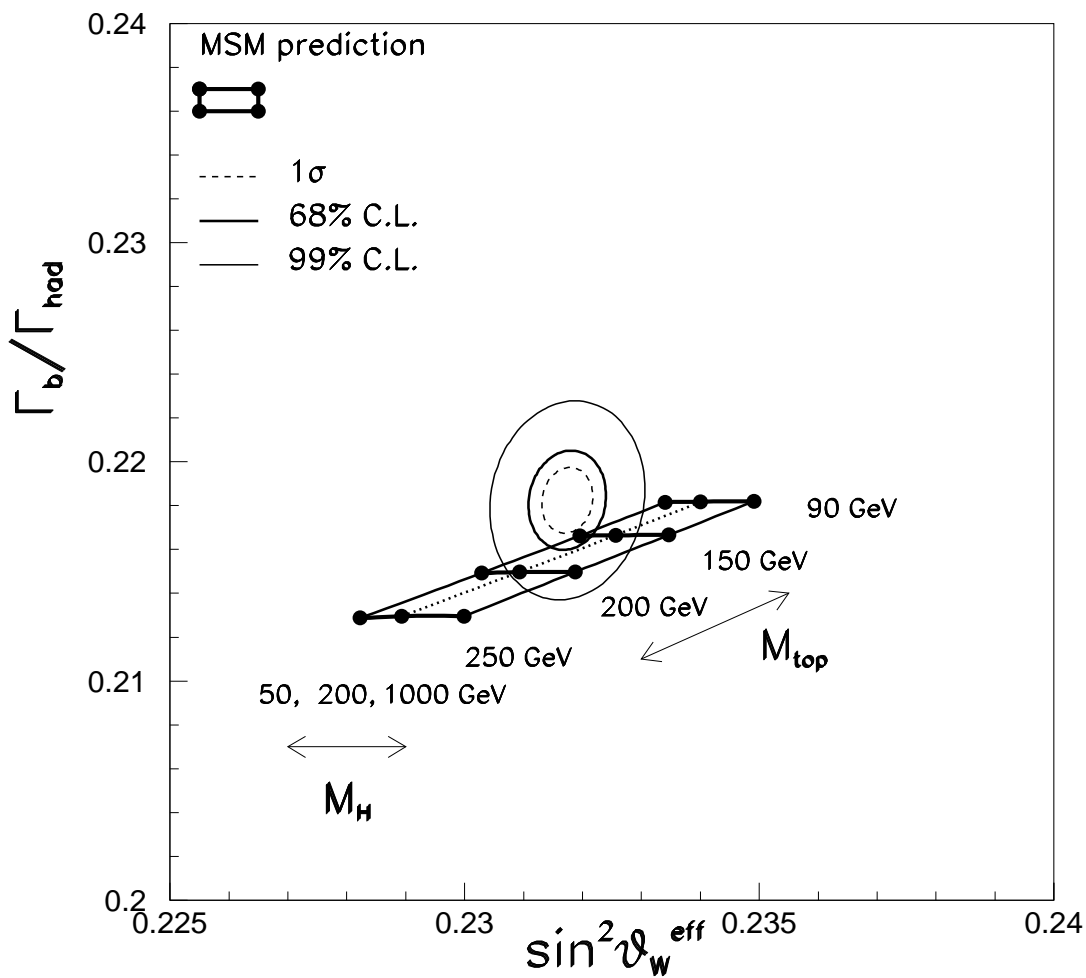

Figure 3: The $R_b$ vs. $\sin^2\theta_W^{eff}$ plane. The intersection of the Standard Model predictions and the fit contour (compare to Figure 2) shows that the preference for low Higgs masses is primarily due to the higher-than-Standard-Model value measured for $R_b$.

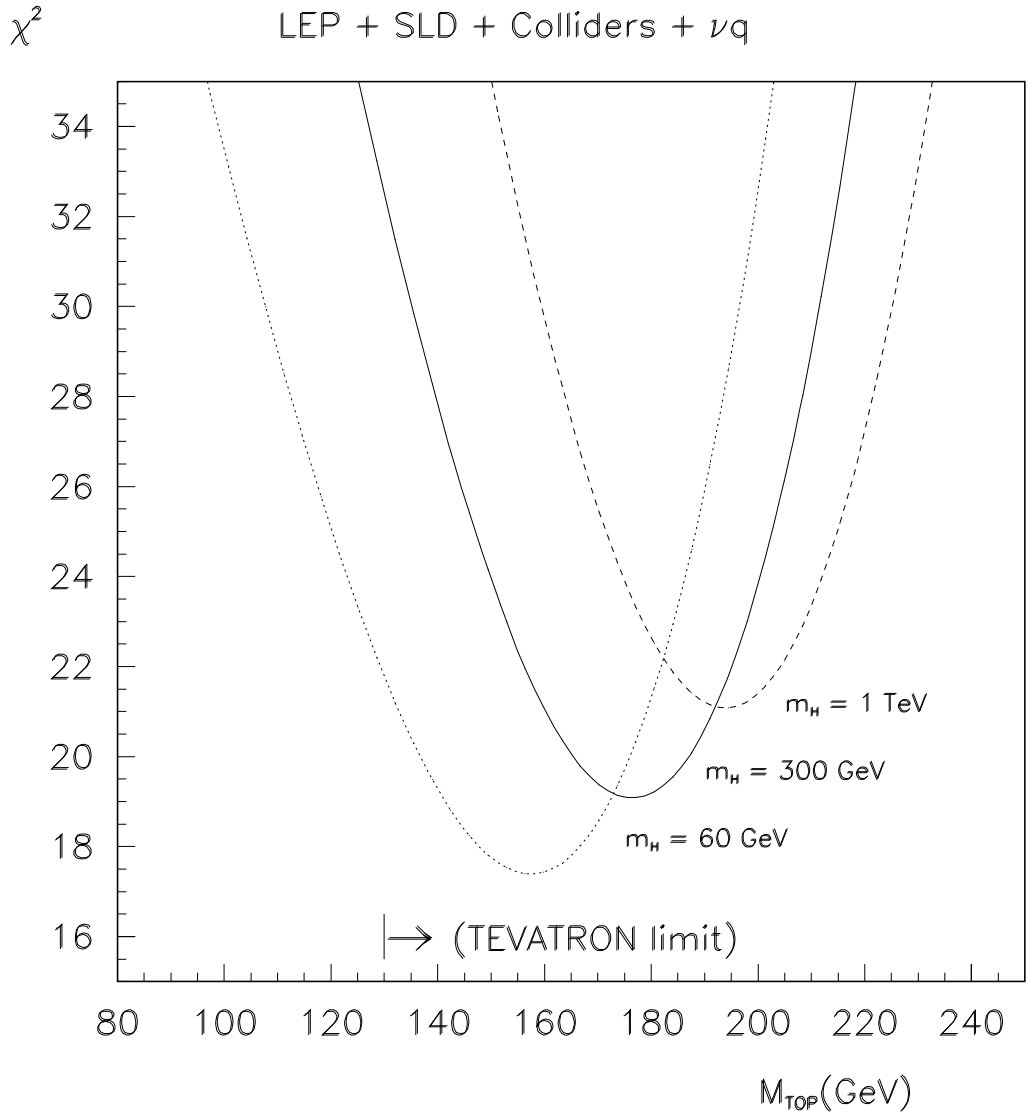

Figure 4: The variation of the $\chi^2$ of the fit to all data as a function of the top mass for three different assumed Higgs masses. The preference for small Higgs masses is not statistically compelling.